\documentclass[12pt,preprint]{aastex}
\usepackage{graphicx-psmin}
\begin{document}
%\bibliographystyle{prsty} % Choose Phys. Rev. style for bibliography

%new commands and definitions

\newcommand{\kms}        {km s$^{-1}$}
\newcommand{\herschel} {\textit{Herschel}}
\newcommand{\water} {H$_2$O}
\newcommand{\hcop} {HCO$^+$}
\newcommand{\chp} {CH$^{+}$}
\newcommand{\den}{${n}_{\mathrm{H}_2}$}
\newcommand{\ncolone}{$N_{\mathrm{CO}}$}
\newcommand{\ncoltwo}{$N(^{13}$CO)}
\newcommand{\tkin}{$T_{\mathrm{kin}}$}
\newcommand{\cmtwo}{cm$^{-2}$}
\newcommand{\cmthree}{cm$^{-3}$}
\newcommand{\coone}{$^{12}$CO}
\newcommand{\cotwo}{$^{13}$CO}
\newcommand{\msun}{M$_{\odot}$}
\newcommand{\lsun}{L$_{\odot}$}
\newcommand{\mic}{$\mu$m}
\newcommand{\htwo}{H$_{2}$}
\newcommand{\lfir} {$L_{\mathrm{FIR}}$}
\newcommand{\chmin}{[$N$(H$_{2})]_{\mathrm{CH}}$}
\newcommand{\tex}{$T_{\mathrm{ex}}$}

\title{First Extragalactic Detection of far-infrared CH rotational lines from the Herschel Space Observatory}

\author{Naseem Rangwala\altaffilmark{1,2},  Philip R. Maloney\altaffilmark{1}, Jason Glenn\altaffilmark{1}, Christine D. Wilson\altaffilmark{3}, Julia Kamenetzky\altaffilmark{1}, Maximilien R. P. Schirm\altaffilmark{3}, Luigi Spinoglio\altaffilmark{4}, Miguel Pereira Santaella\altaffilmark{4}}

\altaffiltext{1}{Center for Astrophysics and Space Astronomy, University of Colorado, 1255 38th street, Boulder, CO 80303} 
\altaffiltext{2}{Visiting Scientist, Space Science and Astrobiology Division, NASA Ames Research Center, Moffet Field, CA 94035} 
\altaffiltext{3}{Dept. of Physics \& Astronomy, McMaster University, Hamilton, Ontario, L8S 4M1, Canada }
\altaffiltext{4}{Istituto di Fisica dello Spazio Interplanetario, INAF, Via del Fosso del Cavaliere 100, I-00133 Roma, Italy}

\begin{abstract}
 We present the first extragalactic detections of several CH rotational transitions in the far-infrared (FIR) in four nearby galaxies: NGC 1068, Arp 220, M 82 and NGC 253 using the \textit{Herschel Space Observatory}. The CH lines in all four galaxies are a factor of 2 - 4 brighter than the adjacent HCN and \hcop\ J = 6-5 lines (also detected in the same spectra). In the star formation dominated galaxies, M 82, NGC 253 and Arp 220, the CH/CO abundance ratio is low ($\sim 10^{-5}$), implying that the CH is primarily arising in diffuse and translucent gas where the chemistry is driven by UV radiation as found in the Milky Way ISM.
In NGC 1068, which has a luminous AGN, the CH/CO ratio is an order of magnitude higher suggesting that CH formation is driven by an X-ray dominated region. Our XDR models show that both the CH and CO abundances in NGC 1068 can be explained by an XDR-driven chemistry for gas densities and molecular hydrogen column densities that are well constrained by the CO observations. We conclude that the CH/CO ratio may a good indicator of the presence of AGN in galaxies.  
We also discuss the feasibility of detecting CH in intermediate- to high-$z$ galaxies with ALMA.

\end{abstract}


\section{Introduction} 
The methylidyne radical CH has been studied extensively at visible wavelengths through its electronic transitions in diffuse Galactic gas \citep{federman97, sheffer08}. From these observations, CH was found to be a powerful tracer of the molecular hydrogen in diffuse and translucent gas. Because CH is a light molecule its ground state rotational transitions lie at submm/FIR wavelengths and are impossible to observe from the ground.  \textit{Herschel} made the first observations of the rotational transitions of CH arising in the far-infrared/sub-millimeter regime in Galactic star forming regions \citep{gerin10b, qin10, bruderer10, naylor10b}. In this work we present the first extragalactic detections of CH in four prototypical galaxies dominated by starbursts or AGN: NGC 1068, Arp 220, M 82 and NGC 253.

CH can be present in high- or low-density gas depending on the formation scenario. If the abundance of ionized carbon is substantial,  CH formation is believed to be initiated by the radiative association of C$^+$ with vibrationally excited molecular hydrogen, \htwo, in the outer layers of photon-dominated regions (PDRs), where the chemistry is dominated by UV radiation. The chemical network forming CH  \citep[described in][]{black73} involves the following reactions:

\begin{eqnarray}\label{chreact}
C^{+} + H_2 \rightarrow  CH_{2}^{+} + h\nu \\
CH^{+}_{2} + H_2 \rightarrow  CH_{3}^{+} + H \nonumber \\
CH^{+}_{2} + e^{-} \rightarrow CH + H \nonumber  \\
CH^{+}_{3} + e^{-} \rightarrow CH + H_2 \nonumber 
\end{eqnarray}

CH can  be formed in high-density gas via  reactions described in Eq.\,[\ref{chreact}] or it can also be produced during \chp\ synthesis in lower density material ($\sim 50$ \cmthree) from MHD shocks \citep{draine86, pineau86}. In Galactic star forming regions, CH is about a factor of 1 -- 3 more abundant than \chp\ \citep{godard12}.  The most efficient reaction forming \chp\ is $C^{+} + H_2 \rightarrow  CH^{+} + H$, which has a high endothermic barrier of 4640 K. This reaction can form \chp\ in a dense and highly illuminated PDR but is inefficient in the cold diffuse ISM. The recent investigations by \citet[][also see \citet{falgarone10}]{godard12, godard09} suggest that \chp\ can be produced in the diffuse ISM via kinetic energy from turbulent dissipation. The \chp\ molecule is also rapidly destroyed at high densities. By comparing CH/\chp\ ratios with other galaxies we can determine whether these molecules are tracing similar environments and densities as the Milky Way (MW) or if their production is influenced by strong starbursts and AGN.

\begin{figure}
\includegraphics[scale=0.3]{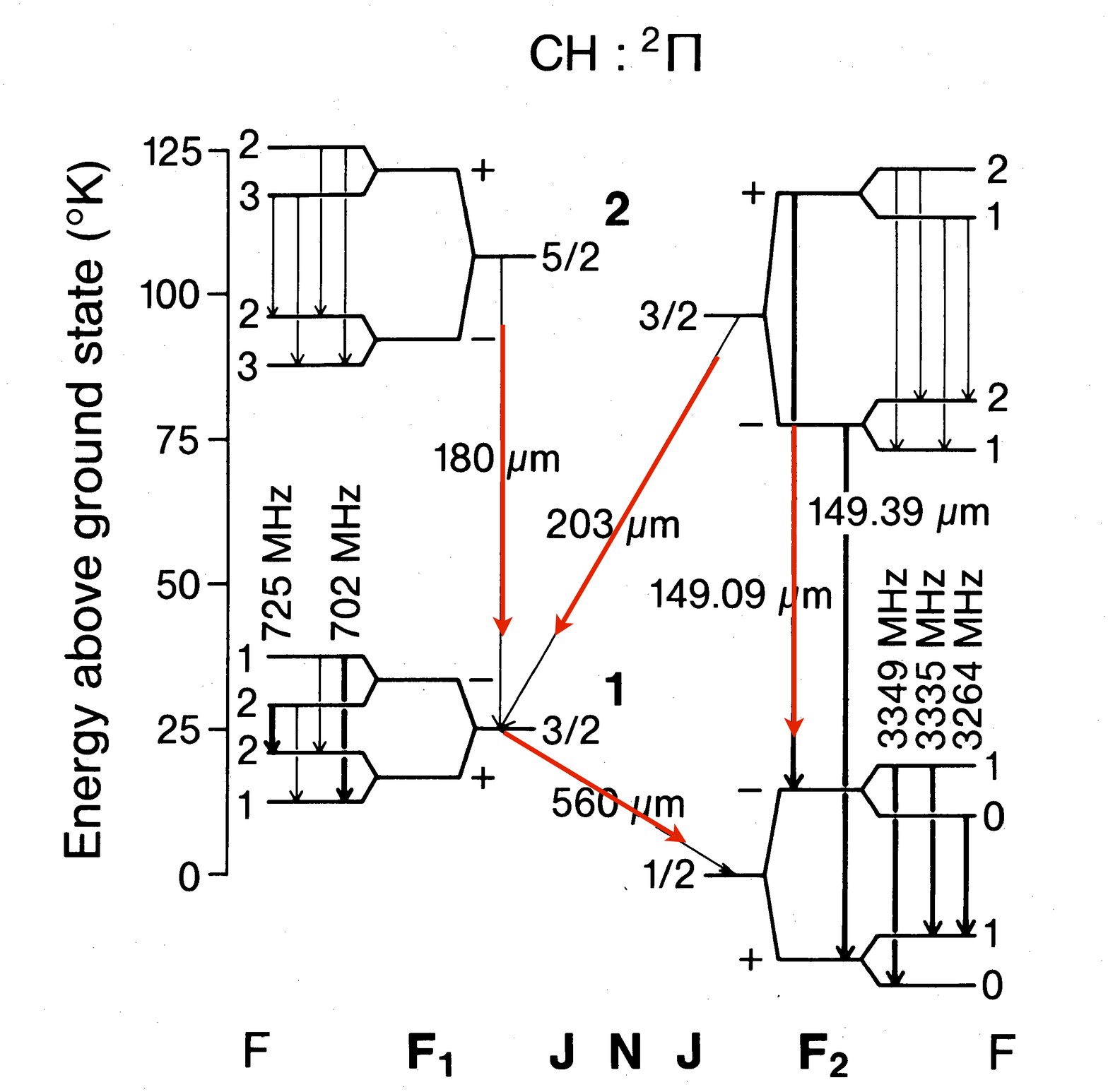}
\caption{The energy level diagram for CH from \citet{stacey87}. The transitions highlighted by the red arrows fall in \textit{Herschel} wavelength bands.}      
\label{chel}
\end{figure}

The CH energy level diagram shown in Figure \ref{chel} is taken from \citet{stacey87}. The rotational lines of CH in the submm/FIR have a characteristic doublet pattern due to lambda doubling: the spin-orbit interaction of the unpaired $\pi$ electron splits the rotational levels (N) into two ladders depending on the relative orientation between the electron's spin and orbital angular momentum vectors. The rotational levels, J, in the individual ladders split into $\Lambda$-doublet states (denoted by +  or $-$) from the relative orientations of the electron's orbital momentum axis and the molecular rotation axis. The magnetic hyperfine interaction further splits the $\Lambda$-doublet states.  The major rotational transitions at 560 \mic, 203 \mic, 180 \mic, and 149 \mic\ (highlighted by the red arrows in Figure \ref{chel}) are accessible with \textit{Herschel}. The 560 \mic\ transition is the easiest to detect by both the SPIRE Fourier Transform Spectrometer (FTS) and Heterodyne Instrument for the Far Infrared (HIFI) because of their high sensitivity at this wavelength. This transition has six hyperfine components  grouped near 532.8 GHz ($1+ \rightarrow 1-, 2+ \rightarrow 1-, \mathrm{and} \, 1+ \rightarrow 0-$) and 536.8 GHz ($2- \rightarrow 1+, 1- \rightarrow 1+, \mathrm{and} \, 1- \rightarrow 0+$), very close in frequency to the HCN and \hcop\ J = 6-5 lines and therefore can only be resolved by the higher resolution of HIFI. 

In this work we present sub-millimeter observations of these CH lines and additionally \chp\ lines from \textit{Herschel} in four prototypical galaxies: Arp 220 (Starburst/AGN), NGC1068 (AGN Seyfert-2), M82 (Starburst/ no AGN) and NGC 253 (starburst/ no AGN). These observations are presented in Section 2. The properties of molecular gas derived from CH are compared with CO observations to investigate where CH is arising in these galaxies and if its formation in these starburst/AGN dominated galaxies differs from the Milky Way (sections 3 and 4). The feasibility of detecting CH in intermediate- to high-redshift galaxies with ALMA is discussed in section 5.

%CH correlates much better with H2 than \chp. 

%-- correlation of CO and CH with N(H2) in the diffuse and dense clouds

\section{Observations and Data Reduction}
\begin{figure}
\includegraphics[scale=0.7, angle=90]{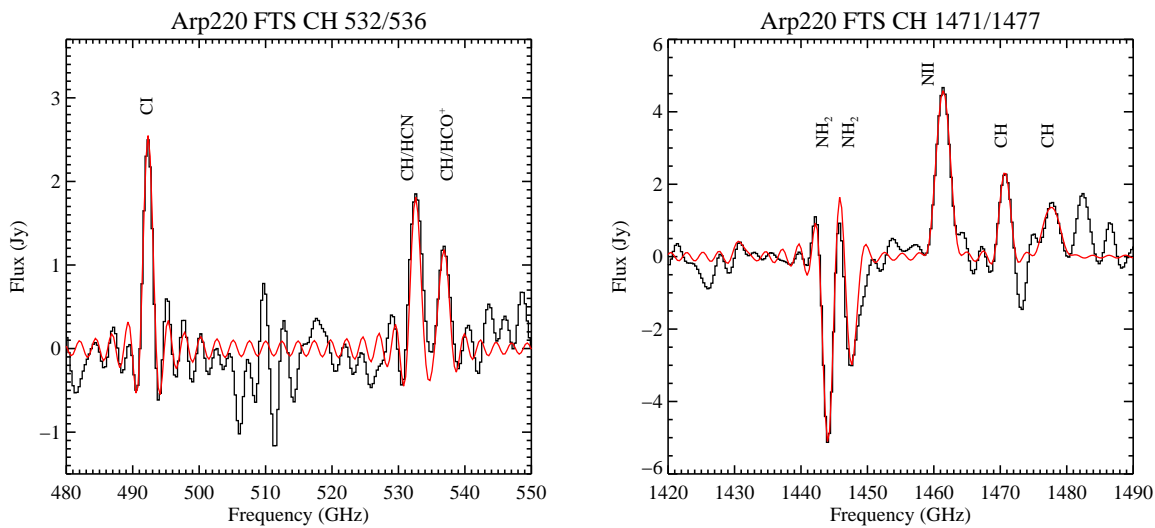}
\caption{CH line detections from the SPIRE-FTS in Arp 220. In the FTS spectra the HCN and HCO+ lines are blended with the CH lines. The neighboring lines are identified for reference.}
\label{chfts}
\end{figure}

The lowest spin-rotational (560 $\mu$m) transition of CH has six hyperfine components grouped at 532.8 and 536.8 GHz. %They were first detected by us 
We first detected them in the SPIRE-FTS spectrum of Arp 220 \citep[][R11 hereafter]{rangwala2011}, as shown in the left panel of Figure \ref{chfts}. In the FTS spectra, they are blended with the HCN and \hcop\ J = 6-5 lines. The presence of CH was inferred from the line center of the blended feature and larger line widths than expected from the FTS resolution. The CH lines are resolved in the HIFI data presented here, unambiguously confirming their detection. We also detected the CH 203 \mic\ transition, consisting of 8 hyperfine components grouped around 1471 and 1477 GHz, in the same FTS spectra, as shown in the right panel of Figure \ref{chfts}.  

We use data from \textit{Herschel}-HIFI to measure the fluxes of the CH, \chp, HCN and \hcop\ lines in the four galaxies. These data come from our Open Time (OT) program (OT1-nrangwal-1) and other programs (both OT and GTO) publicly available on the Herschel Science Archive (HSA). In our OT program we acquired follow-up observations with HIFI to resolve the CH 532/536 GHz lines from the HCN/\hcop\ lines. Additionally, we obtained observations and integrated line fluxes (through Private Communication with J. Fischer and E. Sturm) for the CH 149 \mic\ and CH 180 \mic\ transitions detected in Arp 220 with the PACS spectrometer. The 149 \mic\ line was detected in absorption. Due to blending with a strong water line the 180 \mic\ CH transition is not easily recoverable and therefore its line flux is not reported. The summary of the observations and various programs is provided in Table \ref{obs}. 

\begin{figure}
\includegraphics[scale=0.5]{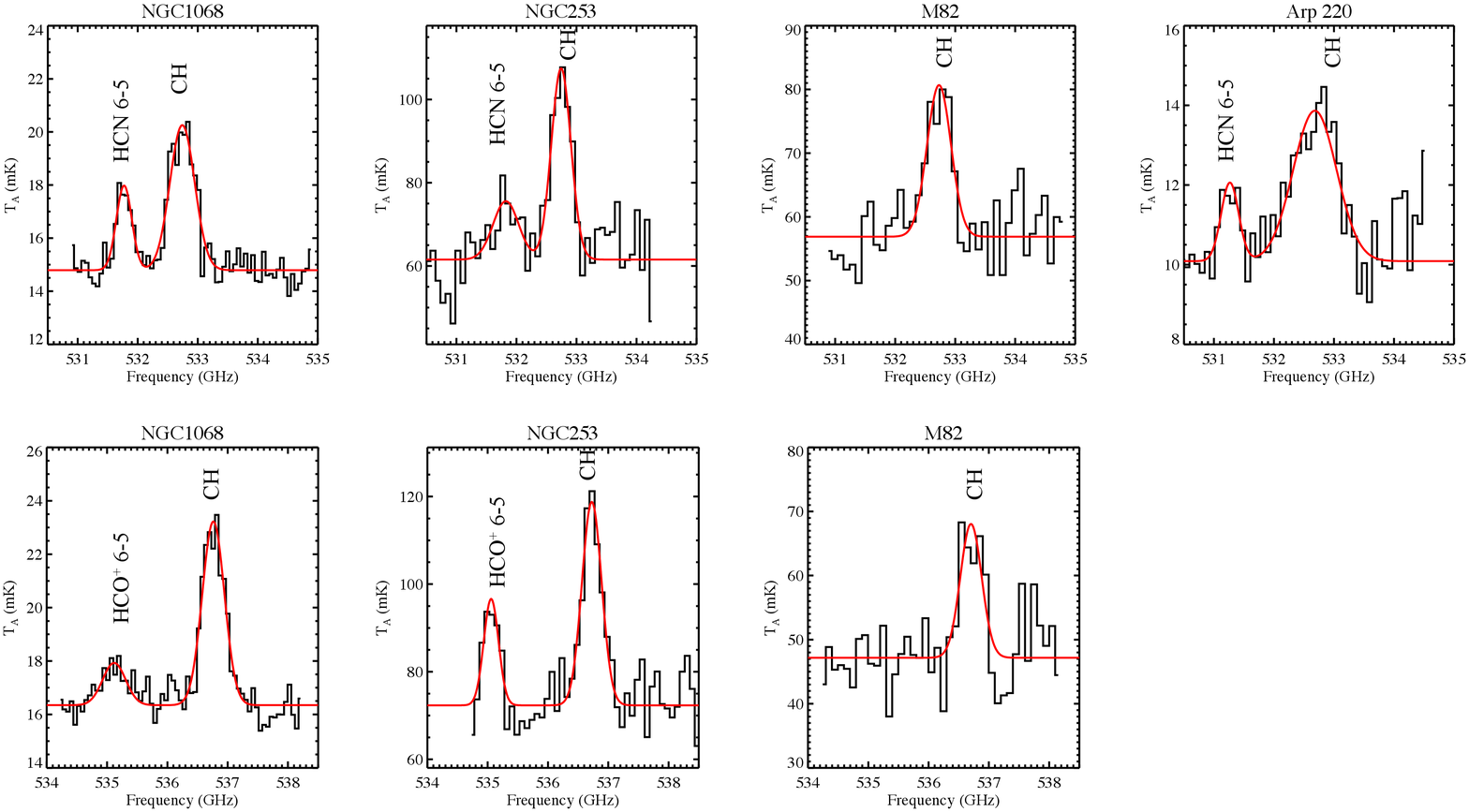}
\includegraphics[scale=0.5]{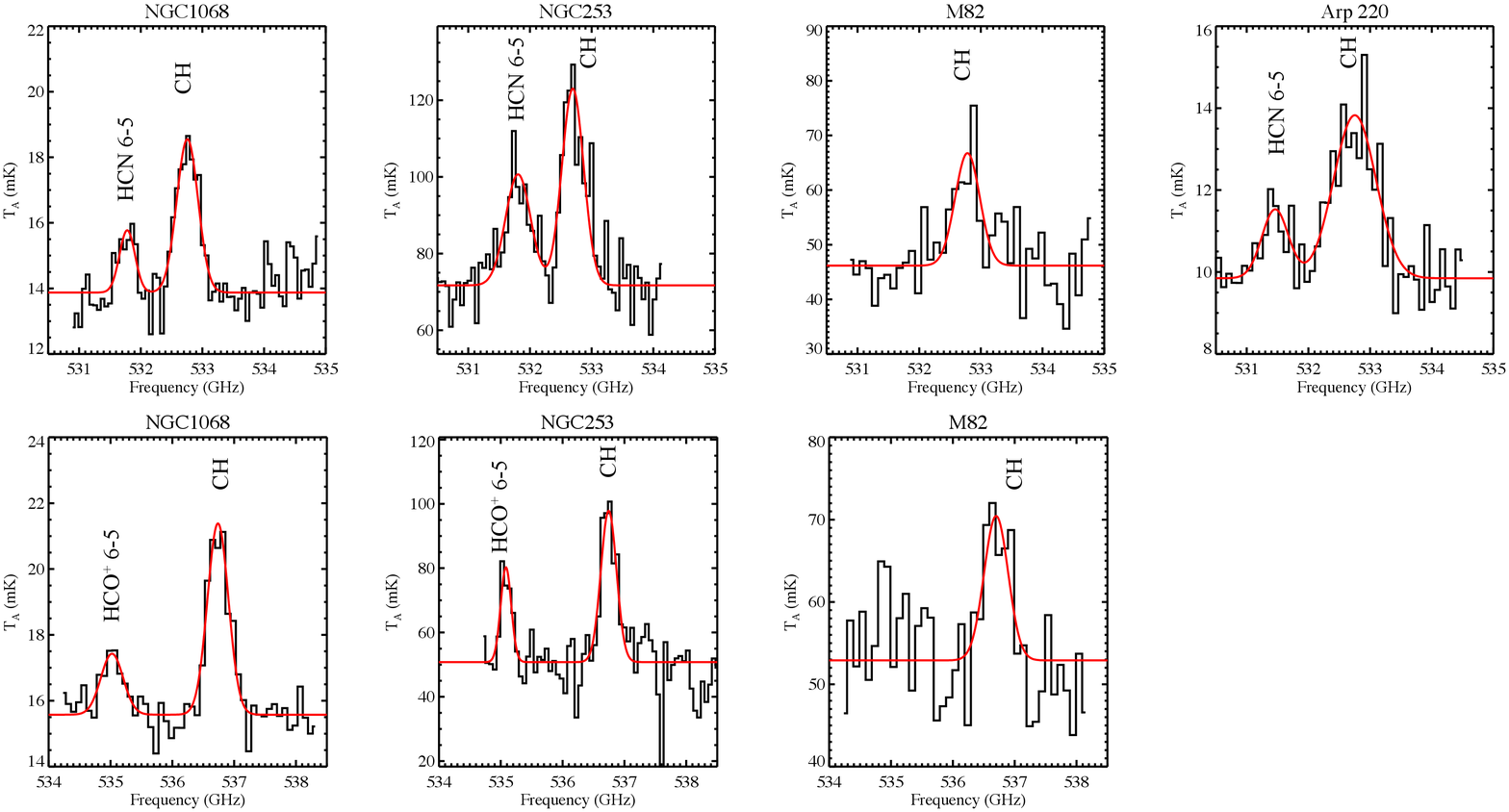}
\caption{CH line detections from HIFI  in four galaxies. Top and bottom panels are for the H- and V- polarization spectra, respectively. The gaussian fits are overplotted in red.}      
\label{hpol}
\end{figure}

For CH, we used the level-2 HIFI spectra that were reduced with HIPE version 9.0. A Gaussian function was fit to the spectral lines to determine the line width (FWHM) and integrated fluxes.  All the spectra were observed in HIFI band-1$a$, in which the baselines are well behaved. We analyzed both V- and H- polarization data separately and report the error-weighted averages of the line widths (FWHM) and integrated fluxes in Table 2. The spectra for the CH 532/536 GHz lines and their Gaussian fits are shown in Figure \ref{hpol} for the H (top panel) and V (bottom panel) polarizations.  The HCN and \hcop\ lines are also detected and are separated from the corresponding CH lines by about 1 and 1.7 GHz, respectively. Table 2 lists the rest frequencies, line width FWHMs, line fluxes in Jy \kms\ and W m$^{-2}$, source or beam size in arcseconds and column densities for CH, HCN and \hcop\ calculated for their respective beam or source size (discussed in Section 3).  

The observations of the \chp\ 835 GHz line are available in the literature and on the Herschel Science Archive for the four galaxies presented in this work. For \chp, we use the line fluxes for NGC 1068 and Arp 220 published in \citet[][S12 hereafter]{spinoglio12} and R11,  respectively. For NGC 253 and M82, the HIFI \chp\ observations are publicly available on HSA. We show these data and their fits in Figure \ref{chp}. The \chp\ appears in absorption in all the galaxies except for NGC 1068, in which it is detected in emission (S12). The equivalent widths and corresponding column densities of \chp\ are listed in Table \ref{chp}. 

\section{Analysis: Column densities of CH and \chp}
The column densities for the transitions of CH, HCN, \hcop\ and \chp\ are calculated assuming that the lines are optically thin. A more accurate way to measure column density would be to use a non-LTE radiative transfer modeling technique \citep[e.g., see, R11,][] {kamenetzky12}. However, this requires observations of several rotational transitions as well as the knowledge of collisional cross-sections (or critical densities) with \htwo. Since neither are available, an optically thin assumption will give a robust lower limit on the column density. 

Under this assumption the column density of an emission line is given by $N_{i} = (F_{i}/(A_{i} h \nu_{i})) \times 4\pi/\Omega$,  where $N_{i}$ is the column density in the upper state, $F_{i}$ is the line flux in [W m$^{-2}$], $A_{i}$ is the Einstein coefficient, $h$ is the Planck constant, $\nu_{i}$ is the frequency of the line in [Hz], and $\Omega$ is the beam or source solid angle in [steradians]. The total column density of a molecule is: $N_\mathrm{mol} = Z(T_\mathrm{ex}) \times (N_{i}/g_{i})\times \mathrm{exp}(-T_{ex}/T_{i})$, where, $T_\mathrm{ex}$ is an excitation temperature in [K], $Z(T_\mathrm{ex})$ is the partition function at $T_\mathrm{ex}$ and $g_{i}$ is the degeneracy of level $i$ \footnote{The values of upper level degeneracies are obtained from the Splatalogue database.}. Except for the CH 149 \mic\ line in Arp 220, all other CH, HCN and \hcop\ lines appear in emission in all of the galaxies and their column densities for individual transitions estimated according to the above relation are listed in the Table 2. 

The column density relation for an optically thin absorption line is given by  $W_{\lambda}/{\lambda} = 8.85 \times 10^{-13} N_{j}\lambda f_{jk}$ \citep{spitzer68}, where $W_{\lambda}$ is the equivalent width, $N_{j}$ is the column density in the $j$th level, and $f$ is the oscillator strength. The unit of the constant in the above equation is [cm]. The column densities of \chp\ lines that appear in absorption are listed in Table \ref{chp}. In the cases of NGC 253 and M 82, there is an overlapping emission line, which may be partially filling in the absorption. Thus, their column densities in Table \ref{chp} are listed as a lower limit.

A complete determination of the total CH column density would require all four major rotational transitions or an estimate of $T_\mathrm{ex}$.   In all of the galaxies except Arp 220 we only detect the 560 \mic\ transition.  Arp 220, being the brightest nearby ULIRG and an important template for high-$z$ galaxies, had much deeper observations from every instrument on \textit{Herschel}. The S/N in the FTS spectra was high enough to detect the 203 \mic\ transition (hyperfine lines grouped around 1471 and 1477 GHz), while the 149 \mic\ (in absorption) and 180 \mic\ lines were detected in the PACS spectra. This makes Arp 220 the only object in which all four transitions have been detected. Even the MW star-forming regions do not have detections of the other three transitions.  We calculated an excitation temperature of $T_\textrm{ex} =  41 \pm 10$ K  in Arp 220 for the relative populations in the upper states of the 560 \mic\ and 203 \mic\ transitions, assuming local thermodynamic equilibrium (LTE). This value is comparable to the kinetic temperature of 50 K derived from low-J transitions of CO and CI, which are tracing primarily the cold gas. 

Using this excitation temperature we calculate a partition function and derive a total CH column density in Arp 220 of about $\sim
1.8 \times 10^{16}$ \cmtwo. This is only a factor of 2 higher than the
column density of the ground state 560 \mic\ transition ($\sim 9 \times
10^{15}$ \cmtwo). If the overall excitation of the CH levels is very subthermal, i.e., if \tex\ for the upper and lower levels of the 560 \mic\ line is small compared to $E_\mathrm{u}$/k, it is possible for most of the column density to be in the ground level, in which case the above estimate would be a significant underestimate of the true column density.
However, this is a very unlikely scenario for several reasons. The upper level of the 203 \mic\ transition lies much further above ground ($\sim100$ K compared with $\sim25$ K) than the upper level of the 560 \mic\ line. We would therefore generally expect that, if the levels are not in LTE, the excitation temperature characterizing these relative level populations (41 K) will be comparable to or \textit{lower} than that of the 560 \mic\ line\footnote{In the limit in which the excitation is very subthermal, i.e., nearly all of the column is in the ground state, it is possible for the excitation temperature of a \textit{transition}, which describes the relative populations of the upper and lower levels, to approach the gas kinetic temperature, even though the excitation temperatures of these levels defined with respect to the \textit{ground} level are much lower, and may approach the temperature of the microwave background (assuming no other radiation background is more important). However, this only occurs when the fractional populations in these levels is negligibly small, meaning that the lines would be unobservable.}. Additionally, in Arp 220, the 149 \mic\ CH line is detected in absorption and therefore gives us a direct measurement of the column density in the ground state. The column density derived from the CH
149 \mic\ absorption line is about $10^{15}$ \cmtwo, almost an order of
magnitude \textit{lower} than that of the 560 \mic\ emission line, confirming that
the column density in the ground state is not large compared to the column density derived from the excited-state transitions seen in emission. In fact, the ground state column density is so low compared to that seen in emission that it is likely that there is some emission partially filling in the absorption.
However, for the true ground state column density to be much larger than our estimate of the total column density, nearly all of the absorption would need to be filled in by emission, while also leaving an absorption column that is coincidentally within an order of magnitude of the column density estimated from the emission lines.
This is a highly improbable scenario. Both the above arguments suggest that in Arp
220, the true CH column density does not significantly differ from the
lower limit derived from the 560 \mic\ emission line.

For the other three galaxies the 203 \mic\ transition is not available
and hence we cannot estimate \tex. However, additional information from
the HCN and \hcop\ J=6-5 lines detected in the same spectra as CH can be
used to put reasonable constraints on \tex. The HCN and HCO+ 6-5 lines
arise from levels much further above the ground ($\sim 90$ K) than the CH
560 \mic\ line and have much larger A-coefficients (by about an order
of magnitude). If the collisional excitation rate coefficients are
similar for the three species, then the critical densities for the HCN and \hcop\ lines will be correspondingly larger than for the CH line, and we would expect that the excitation temperatures of the HCN and \hcop\ J=6-5
levels will be substantially lower than for the CH 560 \mic\ line (unless the line optical depths are large enough that radiative trapping become significant, which is very unlikely for these species). There are no collision rates available in the literature for CH. However, those for OH, which has a very similar electronic structure, are available, and are comparable in magnitude to those for the HCN and \hcop\ lines\footnote{In fact, for this argument to be invalid, the collision rates for CH would have to anomalously small compared to typical collisional rate coefficients.}. For 3 of our 4 galaxies, observations of the HCN J=1-0 transition
are available in the literature. We have used these to calculate an excitation temperature for the populations of J=6 relative to J=1. These numbers range from
13 to 16 K, and (noting again that we expect these numbers to in general
be lower than for the 560 \mic\ transition) argue against a very small
\tex\ for the 560 \mic\ line and therefore against a very large ground
state column density. Also, in all 4 galaxies the CH 560 \mic\ line is
substantially brighter than the HCN/\hcop\ J=6-5 lines. Given that their
abundances (relative to \htwo) are comparable to the abundance of CH
($\sim (2 - 4) \times 10^{-8}$), it would be very surprising for the CH 560
\mic\ line to be much brighter than HCN/HCO+ lines if the CH line has a very low
excitation temperature. On the basis of these arguments, we make the reasonable assumption that the excitation temperature in the other
galaxies is not very different from Arp 220 and use the CH 560 \mic\
line as a proxy for the total CH column density, modulo a factor of 2. Note
that the CO-derived temperatures for the cold molecular gas range from
about 15 -- 50 K in these galaxies \citep[R11, S12,][]{kamenetzky12}.

\section{CH formation, excitation and CH/\chp\ ratio}
\subsection{Comparing CH and CO Column Densities}

The optical observations of CH in the Milky Way show that it primarily arises in the diffuse molecular ($n_\mathrm{H} = 100 - 500 $ \cmthree\ and T = 30 - 100 K) and translucent regions ($n_\mathrm{H} = 500 - 5000 $ \cmthree\ and T = 15 - 50 K) \citep[see review by][]{snow06}. A translucent cloud is a transition region between diffuse and shielded (fully) molecular gas where the incident UV radiation is becoming attenuated and the $C^{+}$ is transitioning to C and CO. In these regions CH is a very reliable tracer of molecular hydrogen. \citet{sheffer08} compiled observations of CH, CO and \htwo\ for many Galactic lines-of-sight. Their results show unambiguously that CO, CH and \htwo\ are tightly correlated. However, as expected there are breaks in these correlations at the boundaries of the diffuse/translucent and translucent/{\bf shielded} regions where the abundances of CH and CO change significantly. The CH/CO ratio is much higher ($10^{-1} - 10^{-3}$) in the translucent regions and starts dropping off dramatically when approaching the shielded and fully molecular regions (CH/CO $\sim 10^{-4}$) where the carbon is almost entirely in CO and the lack of $C^{+}$ halts the formation of CH; see equation (1).  In \citet{sheffer08}, the break in the CH/CO ratio occurs around an \htwo\ column density of $5 \times 10^{20}$ \cmtwo. More recently, \citet{qin10} used HIFI observations of the hyperfine components of the 560 \mic\ line to reach even larger molecular hydrogen column densities: up to $10^{23}$ \cmtwo\ towards the Sagitarius B2 star-forming region. This allowed them to probe the interior of fully molecular clouds. They found that the linear relationship between CH and \htwo\ found in diffuse and translucent clouds does not continue in the denser clouds (or at higher visual extinction). The CH  column density curve flattens around an \htwo\ column density of $\sim 2 \times 10^{21}$ \cmtwo, again implying that CH formation declines in denser shielded regions. To understand where CH arises in our sample of four galaxies we compare CO and CH column densities in this section; the CH formation and its abundance could be different from our Galaxy because the molecular ISM of our sample galaxies are significantly influenced by starburst and AGN activity.

Since the rotational lines of CO are bright and easy to observe in galaxies, its emission has been widely used as a tracer of molecular gas mass in galaxies. By comparing the \htwo\ column densities derived from CH and CO we can also assess whether the CH rotational line transition at 560 \mic\ can be used as a mass tracer in addition to CO. It will be particularly useful to have an additional molecular hydrogen tracer for measuring gas mass and excitation conditions of redshifted star-forming galaxies. This comparison will also allow us to investigate whether CH is arising in the translucent or shielded interiors of fully molecular clouds in these galaxies.

%%%EBB: I WILL CHANGE THIS TO BE CONSISTENT IN TENSE
We estimated the \htwo\ column densities ($N$(\htwo)) from the observations of CH and CO columns densities (N(CH) and N(CO)). The N(CO) for the four galaxies were derived from non-LTE radiative transfer modeling of the observed CO rotational lines from $J = 1-0$ to $J = 13-12$. The mid- to high-J lines were observed with the SPIRE-FTS and the low-J measurements are ground-based observations obtained from the literature. The CO spectral line energy distributions (SLEDs) were modeled using a custom version of RADEX \citep{tak07}, combined with a likelihood code. For a given set of input parameters (\tkin, \den\ and N(CO)), the code starts from an optically thin case to generate the initial guess for the level populations then iterates until a self-consistent solution is achieved such that the optical depths of the lines are stable from one iteration to the next. This code can reliably converge for optical depths up to 100. Model CO SLEDs are produced for a wide range in gas parameters (\tkin, \den, etc.) and compared to the observed SLEDs to generate likelihood distributions of \tkin, \den\ and N(CO). Using multiple CO transitions this code provides a robust determination of the CO column density without requiring an assumption of LTE. The observations, modeling and results for Arp 220, NGC 1068 and M82 are published by R11, S12 (see also \citet{haileydunsheath12}), and \citet{kamenetzky12}. In the cases of the starburst galaxies Arp 220 and M 82, the low-J lines (up to J = 3-2) trace cold molecular gas (\tkin: 15 -- 50 K), which dominates the mass of the molecular gas, whereas the mid- to high-J lines trace warm molecular gas (\tkin: 450 -- 1300 K), which dominates the luminosity of the molecular gas. In NGC 1068, which is a Seyfert-2 galaxy, the high-J CO transitions originate in the compact circumnuclear disk and are almost entirely excited by the X-ray radiation from the central AGN. The results for NGC 253 are in preparation; the same procedure as described above was used to derive the column densities for cold and warm CO.
Note that in each galaxy the same source size was used to estimate both N(CH) and N(CO).

The summed cold and warm N(CO) is converted to $N$(\htwo)  by using a CO/\htwo\ abundance ratio of $1 \times 10^{-4}$ \citep{sheffer08} found in the MW clouds. For converting CH column densities to $N$(\htwo) we use two values of CH/\htwo\ abundance ratio: (a) $3.5 \times 10^{-8}$ derived from UV observations of CH and \htwo\ \citep{sheffer08} and (b) $2 \times 10^{-8}$ derived from the \textit{Herschel}-HIFI observations of the 560 \mic\ CH line in Sgr B2 (M) \citep{qin10}. The latter probes a larger \htwo\ column density (deeper into the cloud and higher densities), up to $1 \times 10^{23}$ \cmtwo. Note that $N$(\htwo) cannot be measured directly in dense clouds and in \citet{qin10} it was derived from H$^{13}$CO$^{+}$ column densities. These two assumed abundances define the N(\htwo) range in Table \ref{H2col}. 

The \htwo\ column densities as derived from CO and CH are in reasonably good agreement for NGC 1068, implying that both CH and CO are arising in the dense molecular gas with a CH/CO ratio of $\sim 10^{-4}$. This is in contrast to the Milky Way where the CH is primarily in the diffuse/translucent gas and its abundance relative to CO drops significantly in the dense regions to less than $10^{-4}$ \citep[see Figure 9 of][]{sheffer08}. This implies that the formation of CH in NGC 1068 is driven by an X-ray Dominated Region (XDR) powered by the luminous AGN. The CH/CO ratios in the other three galaxies are much lower; about $4 \times 10^{-5}$ in Arp220 and $2 \times 10^{-5}$ in M82 and NCG 253. If the formation of CH is as in the MW, we would expect a low CH/CO ratio when observing the total line-of-sight molecular hydrogen column density, because it will be dominated by the shielded fully molecular gas where the local CH abundance is very low. Both M82 and NGC 253 have low CH/CO ratios, suggesting that the CH arises in translucent regions in these galaxies also and that the translucent regions contain about 10\% of the molecular gas mass. In Arp 220 this ratio is also lower but not as low as in M 82 and NGC 253. There are many factors that make the interpretation for Arp 220 more complex. The dust optical depth in Arp 220 is much higher, such that our assumption of optically thin lines could underestimate the CH column density. Additionally, there is strong evidence of a highly obscured AGN in Arp 220 \citep[R11,][]{engel11, downes07, alfonso04}, which could affect the abundance of CH by providing alternative chemical formation pathways. The role of AGN in the formation of CH is discussed in detail in the next section in the context of Arp 220 and NGC 1068. 

In addition to CH/CO we also compare the ratio of CH/\chp\ (listed in Table \ref{chp}) between our sample galaxies and the Milky Way, where the ratio ranges from $1 - 3$ \citep{godard12}. For M82 and NGC 253 this ratio is $\lesssim 4$, consistent with the Galactic value. But in NGC 1068 and Arp 220 this ratio is unusually high, about 23 and 560, respectively. This is because the N(CH) column density increases with the far-infrared luminosity of the galaxies, similar to CO column density, but N(\chp) is more or less constant within the sample. It is possible that the formation of CH and \chp\ in NGC 1068 and Arp 220 is governed by AGN or X-ray dominated regions (XDRs)  and not UV radiation; we discuss this possibility in the next section. 

%CH is primarily thought to be produced in the outer layers of PDRs, which are abundant in C$^{+}$, suggesting lower densities and column densities like in the Milky Way. 

\subsection{CH Production in XDR Models}
To investigate the origin of CH in NGC 1068 and Arp 220, along with the CH/\chp\ ratio discrepancy, we have generated XDR models using an updated version of the code described in \citet{maloney96}. We assume a hard X-ray (1--100 keV) luminosity of $10^{44}$~ergs~s$^{-1}$, as would be expected for an object of Arp 220's luminosity if a large fraction of the bolometric luminosity ($\sim$ \lfir) were produced by an AGN. These models are generated for an object with the properties (such as hydrogen column and X-ray luminosity) of Arp 220 but the results can be applied to other galaxies, as they scale almost linearly with the X-ray luminosity and XDR column density. A power-law index $\alpha = 0.7$ is assumed. Unlike the models in \citet{maloney96}, we assume a sharp lower energy cutoff of the incident spectrum at 1 keV such that everything except for hard X-rays have already been filtered out by intervening gas closer to the AGN. This assumption affects the results only at column densities $\lesssim 10^{22}$ cm$^{-2}$. We assume an XDR column density that ranges from $1\times10^{20} - 3 \times 10^{24}$ \cmtwo; note that this column also attenuates the X-ray flux. The upper limit of this range is based on the column density derived for Arp 220 using CO observations (R11). The distance between the XDR and the X-ray source (AGN) is assumed to be 50 pc.  The physical and chemical state of the gas is calculated using an iterative scheme; radiative transfer of cooling radiation is handled with an escape probability treatment, including the effects of dust trapping of line photons. 
%({\bf Phil - Is this where I should mention that the column density to the X-ray source is not fixed, it varies as the XDR column increases})
%The column density of an XDR changes by the square root of the change in this distance parameter so even though the value of this parameter is not known for NGC 1068 we know how it will effect the XDR column density.

Figure \ref{xdr} shows contours for N(CH) and the N(CH)/N(CH+) ratio generated by our XDR code for a range of densities and column densities of molecular gas. The kinetic temperature of the gas is over plotted in blue contours. The shaded regions encompass the 2-$\sigma$ range of molecular hydrogen column density and gas density derived from CO (listed in Table 3) in Arp 220 and NGC 1068. The hard X-ray photons are capable of penetrating much larger column densities than the UV photons in PDRs, so CH may form through a similar process as described in Equation 1, even deep in  molecular clouds. In addition, however, the radiative association reaction $C + H \rightarrow CH + h\nu$ and the neutral-neutral reaction $C + H_{2} \rightarrow CH + H $, which has a substantial activation barrier, may play a role or even dominate. The last reaction dominates the formation rate in the NGC 1068 portion of Figure 5 (i.e., attenuating columns of about $10^{22}$ \cmtwo), while radiative association with contributions from the sequence of equation (1) are more important for the Arp 220 parameters. \chp\ formation occurs largely through charge exchange with $H^+$ and $C^+$, i.e., $ H^{+} + CH \rightarrow CH^{+} + H$ or else through ion-molecule reactions such as $H_{3}^{+} + C \rightarrow CH^{+} + H_2$. Our models show that the CH abundance is much larger in an XDR compared to the Milky Way value. This is in agreement with Meijerink et al.\,(2007), who also find the CH abundance relative to other molecules to be significantly greater in XDRs compared to PDRs. 
%%%
%%%EBB: Should this read "...Spaans (2007), which also find the CH..."?  Do all their models find this or are there specific models that find this?  I think it should be "which" with the comma; i.e., a non-restrictive clause
%%%

In NGC 1068, both the observed CH column density and  CH/\chp\ ratio can be produced by an XDR at densities $> 10^5$ \cmthree. The constraints on the density and temperature of the gas in NGC 1068 were obtained from non-LTE radiative transfer modeling of high-J CO observations from \textit{Herschel} SPIRE-FTS \citep[S11,][]{haileydunsheath12}. It was found that the high-J lines tracing the molecular gas in the central region ($\sim 4\arcsec$) of NGC 1068 were excited by an AGN and the gas density and kinetic temperature were constrained to lie between $10^{4.5} - 10^{6.5}$ \cmthree\, and 170 -- 570 K, respectively. This shows that both CH and CO observations can be reproduced for similar excitation conditions, suggesting that the CH and CO are most likely coming from the compact nuclear region of NGC 1068 and tracing the same gas. This also explains why the hydrogen columns derived from CH and CO are in good agreement in this case. This is the only object in our sample for which CH formation can be unambiguously explained by an XDR.  This, combined with a large difference ($\sim 1$ order of magnitude) in the relative abundance of [CH/CO] between NGC 1068 and the three starburst galaxies suggests that the CH/CO ratio could potentially serve as an AGN diagnostic. Note that for the above analysis we are using a lower limit on the total CH column density. The total CO column density is robustly determined from non-LTE modeling of multiple CO transitions. Thus, a lower limit on the total CH column density will give a lower limit on the CH/CO ratio. As mentioned in section 3, at a kinetic temperature of $\sim 50$ K (as measured in CO cold molecular gas) the total CH column density is underestimated only by a factor of 2.  In NGC 1068, where the CH appears to be arising in a warmer CND, the CH column density will likely be underestimated by a larger factor driving the CH/CO ratio even higher.

The situation is more complex in Arp 220. R11 found using CO observations that the density of the molecular gas was constrained very tightly around $10^{3}$ \cmthree. Therefore, the molecular gas as traced by CO emission cannot arise in an XDR because at such low gas densities, the high ionization fractions result in rapid chemical destruction of the CO. A non-ionizing source, such as mechanical energy from stellar winds and supernovae, could explain the total observed CO luminosity in Arp 220.  
 If CH and CO originate from the same molecular component then the CH column density and formation also cannot be explained by an XDR with a density of $10^{3}$.  Figure 5 shows that in the XDR models the expected column density of CH would be an order of magnitude higher than observed given the density and column density constraints of Arp 220.
 A lower CH/CO ratio implies that the CH in Arp 220 is arising in the translucent regions similarly to M82 and NGC 253. We note that the \htwo\ column densities derived from CH and CO differ by a factor between 3 -- 6, which would imply that 15\% - 30\% of the molecular mass is in the diffuse-translucent phase. This mass fraction is higher than expected but the assumed CH/\htwo\ abundance used in converting the CH column density into an \htwo\ column density has a large uncertainty.
 
  %However, the area filling factor for diffuse CH gas could well be much larger than CO, which is originating in more compact regions. A larger filling factor would lower the inferred CH column density and also the mass fraction of the inferred diffuse/translucent gas in Arp 220.
 
The very high CH/\chp\ ratio in Arp 220, compared to the MW and other galaxies, also cannot be reproduced by an XDR at low gas densities. This leads us to consider another possibility - the unresolved \chp\ line observed in Arp 220 is also a mixture of emission and absorption, as seen in the CH 149 \mic\ line, which would make its true column density higher and hence the actual CH/\chp\ ratio could be as much as an order of magnitude lower than the observed value of 560.  This is a very reasonable possibility for Arp 220 because other molecular lines have been observed in emission as well as absorption (e.g.,  HCN). 

In conclusion, in NGC 1068 the formation of CH and CO in the molecular gas is consistent with an XDR suggesting that the CH/CO ratio could potentially be a powerful AGN diagnostic tool. The other three galaxies have much smaller CH/CO ratios suggesting that CH is tracing the molecular gas in translucent regions and its formation mechanisms are similar to the MW. 

\subsection{CH Excitation: Radiation versus Collisions}
In the Milky Way, the CH rotational lines tracing the diffuse ISM are believed to be radiatively excited by the cosmic microwave background radiation \citep{gerin10b, bruderer10}. In our sample of galaxies, the dust radiation field can be much more intense than in the Milky Way. For example, in Arp 220, the dust optical depths are high and the dust temperature is $\sim 67$~K. At this temperature, the radiative pumping rate ($B\,J_{\nu}$) is $\sim 10^{-4} \, \mathrm{s}^{-1}$ assuming a blackbody and no geometric dilution. In comparison, the collision rate of CH (with \htwo) is $\sim 10^{-7}\, \mathrm{s}^{-1}$ for a gas density of $\sim 10^{3}$ \cmthree\ (determined from  CO modeling in our previous work) and an assumed collision cross-section of $\sim 10^{-15}\, \mathrm{cm}^{2}$ \citep[e.g.,][]{bertojo76}; the collision cross-sections for CH are not available in the literature and the value used here is an approximation.  Plausible corrections for optical depth and geometric dilution are unlikely to substantially reduce the large gap between the radiative and collisional excitation rates. Thus, we believe that the CH is most likely excited by radiation in Arp 220. Similar gas densities ($\sim 10^{3}$ \cmthree) in M82 and NGC 253 and the large difference in the radiative and collision rates suggest that CH is most likely also radiatively excited in these galaxies. In NGC 1068, the gas density in the CND is much higher ($\sim 10^{6}\, \mathrm{cm}^{3}$) and the dust optical depths are much lower in Arp 220. The optical depth is lower in M82 and NGC 253 also, but not likely enough to disfavor radiative excitation. For optically thin dust emission, the background radiation field is dominated by the cosmic microwave background radiation, and collisional excitation/de-excitation are more important than radiation excitation.
%%%
%%%EBB: No Numbers on this last statement??
%%%

\section{Detecting CH in High-$z$ Galaxies}
Our observations of CH in four nearby galaxies suggest that the CH/CO ratio can potentially be used as an AGN diagnostic.  This is also supported by the models of Meijerink et al.\,(2007), in which the CH abundance is significantly enhanced relative to other molecules in interstellar media of galaxies with AGN. However, more observations are needed to establish the CH/CO ratio as an AGN diagnostic. This would be extremely useful for probing AGN activity in high-$z$ galaxies, for which determining the presence of an AGN and its influence on the excitation of the molecular ISM is challenging. In addition, determining whether CH is coming from diffuse or dense gas will allow us to determine mass fraction and excitation of the diffuse and dense molecular gas in high-z galaxies. With the ALMA observatory CH observations can be made for galaxies with redshift $z \gtrsim 0.1$. ALMA's large bandwidth will allow simultaneous measurements of HCN and \hcop\ lines, enabling the determination of other line ratios, such as CH/HCN, CH/\hcop\ and HCN/\hcop, which can also provide additional AGN diagnostics (Meijerink et al. 2007).   Furthermore, for galaxies at $z = 3 - 5$, multiple CH rotational transitions are accessible with ALMA, increasing the accuracy of the measurements of molecular mass and excitation conditions, such as kinetic temperature and gas density. However, theoretical calculations of CH collisional cross-sections are needed for non-LTE radiative transfer modeling.  A CH and CO line survey for an adequate sample can be easily accomplished with only a few hours of integration time with ALMA, which will allow us to determine whether (a) CH is arising in the diffuse, translucent or dense molecular gas, (b) it is tracing different temperature and density gas compared to CO, and (c) it can be used as a potential AGN-diagnostic line. 

%AI (a) I still need to add a line about the need for CH collision cross-section, 
 %{\bf the abundance in XDR models are relative to H and not \htwo\ - check if I fixed this }
\newpage

%\section{Summary}
\acknowledgements

We are grateful to Eckhard Sturm, Jacqueline Fischer, Eduardo Gonzalez-Alfonso and Will Lewitus for providing us reduced PACS data and line fluxes for the CH 149 and 180 \mic\ lines in Arp 220. We thank Eric Burgh for important discussions about CH, CO and \htwo\ correlations in the Galactic ISM. We also thank Edith Falgarone for useful discussions about the formation of \chp\ in the ISM.

\begin{figure}
\includegraphics[scale=0.7]{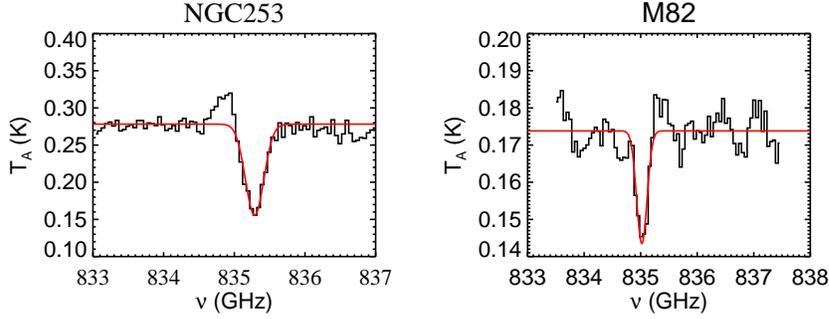}
\caption{HIFI observations of \chp\ in NGC 253 and M 82.}
\label{chp}
\end{figure}

\begin{figure}
\includegraphics[scale=0.5]{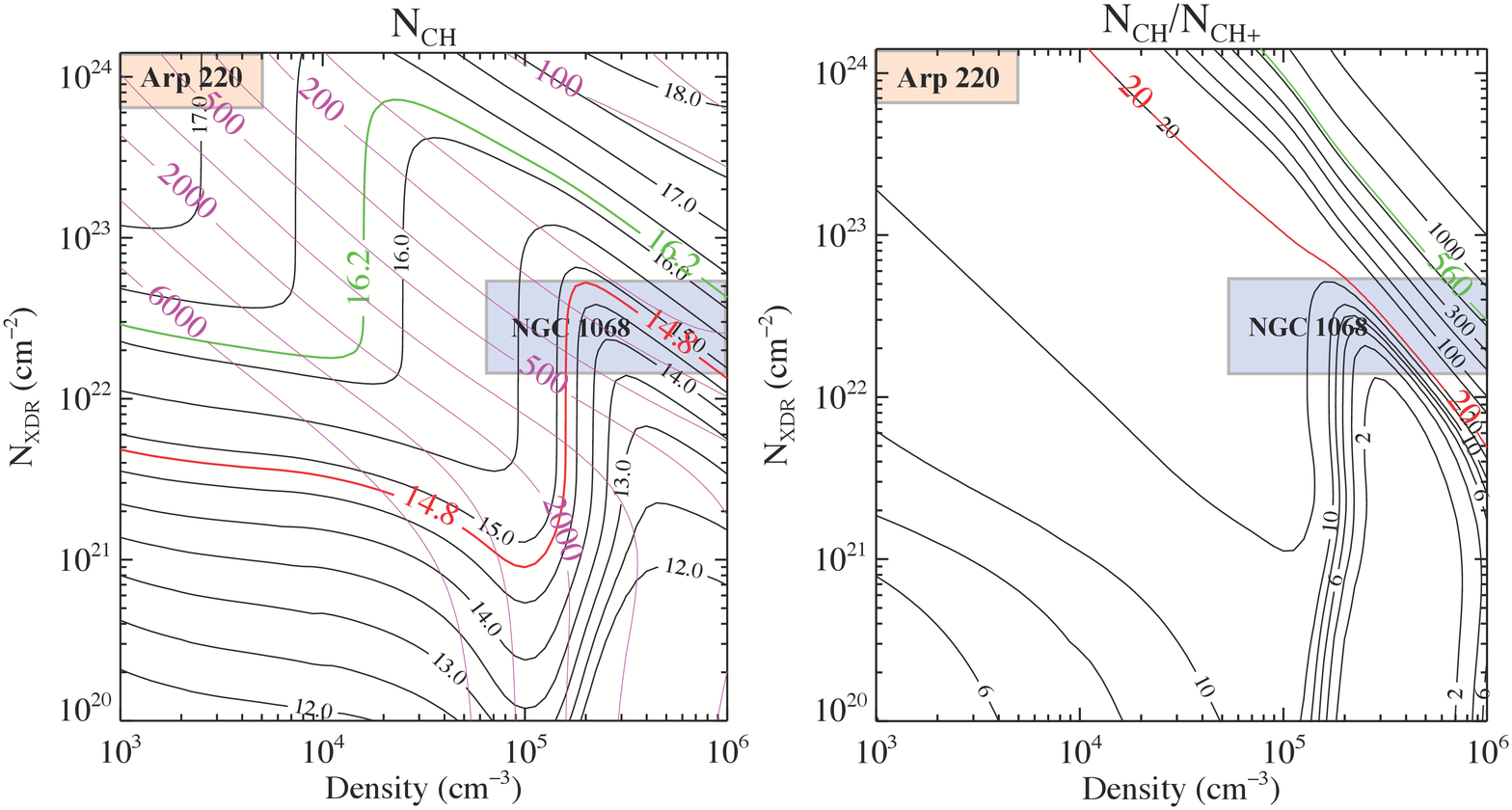}
\caption{Left: CH column densities (in log units) in black solid lines produced from XDR models. The kinetic temperatures are shown by blue contours. Right: Ratio of CH to \chp\ column densities. The shaded regions in both panels show constraints for molecular hydrogen column density and gas density for Arp 220 and NGC 1068 derived from CO observations in \citet{rangwala2011} and \citet{haileydunsheath12}, respectively. The red and green lines are the observed values for CH column density and CH/\chp\ ratio in NGC 1068 and Arp 220, respectively.}
\label{xdr}
\end{figure}

%\begin{figure}
%\includegraphics[scale=0.7]{f6.ps}
%\caption{The CH 560 \mic\ line flux scaled from Arp 220 for a few high-z lensed galaxies as shown by solid lines. ALMA integration times are plotted for S/N of 10 on the integrated flux (See Section 5 for the details involved in calculating these integration times).  The top axis shows the frequency range for ALMA over which this CH lines can be observed.}
%\label{alma}
%\end{figure}
%
\newpage

%\documentclass{aastex}
%\newcommand{\hcop} {HCO$^+$}
%\begin{document}
\begin{deluxetable}{llllll}
\tablecaption{Observations of CH}
%\tabletypesize{\scriptsize}
\tablehead{\colhead{Galaxy}& \colhead{Lines} & \colhead{OBSID} & \colhead{OD} &  \colhead{t$_\mathrm{int}$ (s)} & \colhead{Program ID}
}

\startdata
        M 82 & CH 560, HCN 6-5 & 1342232963 & 925 & 94 & OT1 (N. Rangwala) \\
        M 82 & CH 560, \hcop\ 6-5& 1342232964 & 925 & 82 & OT1 (N. Rangwala)\\
        M82 & CH$^+$ 359 & 1342246037 & 1106 & 1995 & OT2 (E. Falgarone) \\
        NGC 1068 & CH 560, HCN 6-5 & 1342247837 & 1154 & 6891 & OT1 (N. Rangwala) \\
        NGC 1068 & CH 560, \hcop\ 6-5 & 1342237609 & 980 & 7137 & OT1 (N. Rangwala)\\
        NGC 253 	& CH 560, HCN 6-5 & 1342210772 & 568 & 64 &  KPGT (R. Guesten) \\
        NGC 253 & CH 560, \hcop\ 6-5 & 1342210773 & 568 & 40 & KPGT (R. Guesten) \\
        NGC 253 & CH$^{+}$ 359 & 1342212138 & 595 & 348 & KPGT (R. Guesten) \\
        Arp 220 & CH 560, HCN 6-5 & 1342262569 & 1357 & 3453 & OT2 (P. Maloney)\\
        Arp 220-FTS & CH 560 & 1342190674 & 275 & 10445 & KPGT (C. Wilson) \\
        Arp 220-FTS & CH 203 &1342190674 & 275 & 10445 & KPGT (C. Wilson) \\
        Arp 220-PACS & CH 149 & 1342191305& 289 & 3447 & KPGT (E. Strum) \\
        Arp 220-PACS & CH 180 & 1342191309& 289 & 3505 & KPGT (E. Strum) \\
        
        \enddata
\label{obs}
\tablecomments{The CH$^{+}$ observations  for Arp 220 and NGC 1068 are published in \citet{rangwala2011} and \citet{spinoglio12}}
\end{deluxetable}
%end{document}

%\documentclass{12pt,preprint]{aastex}
%\begin{document}
\begin{deluxetable}{llllllll}
\tablecaption{Line Fluxes and Column Densities}
\tabletypesize{\scriptsize}

\tablehead{\colhead{Galaxy}& \colhead{Molecule/} & \colhead{$\nu_{\textrm{rest}}$} & \colhead{Line Width} & \colhead{Flux} & \colhead{Flux} &
\colhead{Size\tablenotemark{a}} & \colhead{$N\tablenotemark{c}$} \\
\colhead{} & \colhead{Transition}& \colhead{(GHz)} & \colhead{km s$^{-1}$} & \colhead{(Jy km s$^{-1}$)} & \colhead{($\times\, 10^{-17}$ W m$^{-2}$)} &
\colhead{($\prime\prime$)} & \colhead{($ $cm$^{-2}$)}}

\startdata    
         M82  &     CH 560  &    532.730  &     265  $\pm$    18  &  3414  $\pm$     170  &      6.10  $\pm$      0.30  &      43.5  &      1.38E+13  \\
         M82  &     CH 560 &    536.760  &     240  $\pm$    18  &  2829  $\pm$     267  &      5.10  $\pm$      0.48  &      43.5  &      1.12E+13  \\
     NGC1068  &     CH 560  &    532.730  &     255  $\pm$     7  &   635  $\pm$      22  &      1.10  $\pm$      0.04  &       4.0  &      2.95E+14  \\
     NGC1068  &     CH 560  &    536.760  &     235  $\pm$     5  &   725  $\pm$      21  &      1.30  $\pm$      0.04  &       4.0  &      3.38E+14  \\
     NGC1068  &    HCN 6-5  &    531.716  &     170  $\pm$     9  &   211  $\pm$      16  &      0.37  $\pm$      0.03  &       4.0  &      6.19E+12  \\
     NGC1068  &   HCO$^{+}$ 6-5  &    535.062  &     245  $\pm$    21  &   202  $\pm$      13  &      0.36  $\pm$      0.02  &       4.0  &      3.46E+12  \\
      NGC253  &     CH 560  &    532.730  &     225  $\pm$     8  &  6052  $\pm$     242  &     11.00  $\pm$      0.43  &      43.5  &      2.49E+13  \\
      NGC253  &     CH 560  &    536.760  &     190  $\pm$     8  &  4928  $\pm$     235  &      8.80  $\pm$      0.42  &      43.5  &      1.94E+13  \\
      NGC253  &    HCN 6-5  &    531.716  &     275  $\pm$    18  &  3187  $\pm$     259  &      5.70  $\pm$      0.46  &      43.5  &      8.06E+11  \\
      NGC253  &   HCO$^{+}$ 6-5  &    535.062  &     127  $\pm$     8  &  1972  $\pm$     147  &      3.50  $\pm$      0.26  &      43.5  &      2.84E+11  \\
      Arp220  &     CH 560  &    532.730  &     470  $\pm$    20 &   986  $\pm$      54  &      1.80  $\pm$      0.09  &       1.3  &      4.57E+15  \\
      Arp220  &    HCN 6-5  &    531.716  &     227  $\pm$    25  &   230  $\pm$      32  &      0.41  $\pm$      0.06  &       1.3  &      6.49E+13  \\
  Arp220-FTS  &     CH 560\tablenotemark{b}  &    536.760  &       ... &  1025  $\pm$     114  &      1.80  $\pm$      0.20  &       1.3  &      4.43E+15  \\
  Arp220-FTS  &    CH 203  &   1470.660  &      ...    &  1506  $\pm$     482  &      7.40  $\pm$      2.40  &       1.3  &      1.13E+15  \\
  Arp220-FTS  &    CH 203  &   1477.620  &       ...       &  1507  $\pm$     500  &      7.40  $\pm$      2.50  &       1.3  &      1.15E+15  \\
  Arp220-PACS & CH 149 & 2010.45 & 272  &  -5268 $\pm$ 260 & -34.7 $\pm$ 1.7 &  1.3 & 1.00E+15\\
        \enddata
\label{data}
\tablenotetext{a}{The Size column refers to FTS beam size for M 82 and NGC 253, and source size for Arp 220 and NGC 1068.}
\tablenotetext{b}{blended with \hcop\ 6-5}
\tablenotetext{c}{The column densities listed here are lower limits as they calculated under the assumption of optically thin lines. Uncertainties on $N$ are directly proportional to the uncertainties in the line flux}
\end{deluxetable}
%\end{document}

\begin{deluxetable}{llrcl}
\tablecaption{CH$^{+}$ column densities}
%\tabletypesize{\scriptsize}

\tablehead{\colhead{Galaxy}  & \colhead{$W_{\lambda}$} &  \colhead{$N_{\mathrm{mol}}$} & \colhead{$X_\mathrm{CH_{560}}/X_\mathrm{CH^{+}}$} & \colhead{Reference\tablenotemark{a}} \\
\colhead{} &  \colhead{$\mu$m} & \colhead{(cm$^{-2}$)} & \colhead{}}

\startdata
Arp 220 & 0.19 & $1.6 \times 10^{13}$ & 560 & Rangwala et al. (2011)\\
NGC 1068 & -- & $3.0 \times 10^{13}$ & 23 & Spinoglio et al. (2012)\\
NGC 253 & 0.14 & $\gtrsim 1.2 \times 10^{13}$ & $\lesssim 4$ & This Work \\
M 82 & 0.074 & $\gtrsim 6.3 \times 10^{12}$ & $\lesssim 4$ & This Work\\

\enddata
\label{chp}
\tablenotetext{a}{References for \chp\ data}
\end{deluxetable}
%\documentclass{aastex}
%\begin{document}
\begin{deluxetable}{llllll}
%\tabletypesize{\scriptsize}
\tablecaption{\htwo\ column densities from CO and CH}
\tablehead{\colhead{Galaxy}& \colhead{$L_\mathrm{FIR}$}& \colhead{X[CH/CO]\tablenotemark{a}}&  \colhead{[N(H$_{2})$]$_{\mathrm{CH}}$\tablenotemark{b}} & \colhead{[N(H$_{2}$)]$_{\mathrm{CO}}$} & \colhead{References\tablenotemark{c}}\\
\colhead{} & \colhead{($L_\mathrm{sun}$)} & \colhead{} &  \colhead{(cm$^{-2}$)}& \colhead{(cm$^{-2}$)} & \colhead{} }

\startdata
 Arp 220    &  $1.8 \times 10^{12}$ & $4.5 \times 10^{-5}$ &  $ (4.0 - 7.1) \times  10^{23}$    &  $2.0^{+0.0}_{-1.2} \times 10^{24}$ &    Rangwala et al. (2011) \\
  NGC 1068 &  $2 \times 10^{11}$  &  $ 1\times 10^{-4}$ & $(1.8 - 3.2) \times  10^{22}$  &  $4.4^{+0.6}_{-1.9} \times 10^{22}$  &   Spinoglio et al. (2012)\\
  NGC 253    &  $2 \times 10^{10}$ & $ 2 \times 10^{-5}$ & $(1.3 - 2.2) \times  10^{21}$   & $2.3^{+3.7}_{-1.9} \times 10^{22} $   & This Work\\
  M 82   &    $5.6 \times 10^{10}$     &  $2 \times 10^{-5}$ &  $(0.7 - 1.3) \times  10^{21}$  & $1.2^{+1.8}_{-0.2} \times 10^{22}$    &    This Work \\
\enddata
\tablenotetext{a}{Derived using the CH 560 \mic\ line}
\tablenotetext{b}{See text in Section 4.1 for explanation of ranges}
\tablenotetext{c}{References for CO measurements}

%\tablenotetext{a}{For Arp220, the source size used to calculate the column densities was derived from dust SED modeling in Rangwala et al. (2011) and was also used to derive the column density of warm CO gas. The source size for CO (1-0) line is a factor of $\sim 1.6$ larger.}
\label{H2col}
\end{deluxetable}
%\end{document}

%\begin{deluxetable}{lllrclll}
%\tablecaption{CH$^{+}$ column densities}
%\tabletypesize{\scriptsize}
%
%\tablehead{\colhead{Galaxy}  & \colhead{$L_\mathrm{FIR}$} & \colhead{$W_{\lambda}$} &  \colhead{$N_{\mathrm{mol}}$} & \colhead{$X_\mathrm{CH_{560}}/X_\mathrm{CH^{+}}$} & \colhead{$\tau_\mathrm{CO5-4}$} & \colhead{$n_\mathrm{cold}$} &  \colhead{$n_\mathrm{warm}$}\\
%\colhead{} & \colhead{$L_\mathrm{sun}$} &  \colhead{$\mu$m} & \colhead{(cm$^{-2}$)} & \colhead{} & \colhead{} & \colhead{\cmthree} & \colhead{\cmthree}}
%
%\startdata
%Arp 220 & $1.8 \times 10^{12}$ & 0.19 &  $1.6 \times 10^{13}$ & 560 & 3.2 & $10^3$ & $10^{3}$ \\
%NGC 1068 &$2 \times 10^{11}$ & -- &   $3.0 \times 10^{13}$ & 23 & 2.3 & $10^3$ & $10^{4.5}$\\
%NGC 253 & $2 \times 10^{10}$ 0.14 & & $1.2 \times 10^{13}$ & $\lesssim 4$ & 0.2 & $10^{3}$ & $10^{3.6}$ \\
%M 82 & $5.6 \times 10^{10}$  & 0.074 & $> 6.3 \times 10^{12}$ & $\lesssim 4$ & 0.25 & $10^{3}$ & $10^{4}$\\
%\enddata
%
%\tablecomments{(a) The error/range in cold gas densities for M82 and NGC 253 is about 2 dex. The warm gas density is more tightly constraint. In NGC 1068 cold and warm should be replaced with extended disk and compact CND. }
%
% \end{deluxetable}
%
%\bibliographystyle{apj}
%\bibliography{vngs}
\end{document}